\documentclass[aip,graphicx]{revtex4-1}
\usepackage{graphicx}
\usepackage{color}

\draft % marks overfull lines with a black rule on the right

\begin{document}

%\preprint{}

\title{Ultrasensitive, universal single-ion nanodetector}


\author{Namita Narendra}
\affiliation{Elmore Family School of Electrical and Computer Engineering, Purdue University, West Lafayette, Indiana 47907, United States}
\email{nnarendr@purdue.edu}

\author{Tillmann Kubis}
\affiliation{Elmore Family School of Electrical and Computer Engineering, Purdue University, West Lafayette, Indiana 47907, United States}
\email{tkubis@purdue.edu}

\date{\today}

\begin{abstract}
In this paper, a carbon nanotube (CNT) based single-ion detector is proposed and its performance is evaluated with atomistic quantum transport models.
The sensor can detect any ion type without molecule-specific functionalization and allows for continuous real-time ion monitoring. 
A single ion temporarily changes the operating principle of the sensor's CNT field-effect transistor into a resonant tunneling diode.
The concrete device example of this paper showed a source-drain current increase of 5 orders of magnitude induced by a single ion.
\end{abstract}

%\pacs{}% insert suggested PACS numbers in braces on next line

\maketitle


%Traditional ion sensors require many ions to produce detectable current \cite{bobacka2008potentiometric} which makes it challenging to detect low concentration ions.

%\subsection*{New concepts}
%This work introduces a fundamentally new mechanism for ion detection at the nanoscale. Using a Carbon nanotube (CNT) based  field-effect transistor, we show that the presence of a single ion in the CNT channel transforms the CNT device operation into that of a resonant tunneling diode. This mechanism enables a single ion to produce a large modulation in current, allowing detection in the single ion regime. Unlike conventional ion sensors based on functionalization, the proposed device operates through electrostatic coupling between the ion and the CNT and thus provides species agnostic non-destructive ion detection.

\subsection{Introduction}
Single-ion sensing is essential for a range of applications such as the detection of picomolar levels of biomarkers \cite{visser2018continuous,raveendran2020rational,zou2024ion}, the detection of traces of heavy metal ions~\cite{shtenberg2015detection,si2024research}, or nanoscale quantum applications~\cite{yoo2025trapped,bonus2025ultrasensitive}.
Scaling down of the ion sensors to sizes comparable to those of molecules has increased sensitivity and made single-ion sensing feasible~\cite{nakajima2016application,gooding2016single,baaske2016optical,bhardwaj2025microtomy}. 
State-of-the-art ion nanosensors, such as nanopores~\cite{miles2013single, roozbahani2020nanopore, graf2019transverse} and nanowire transistors~\cite{hu2021ion, li2016direct}, have drawn attention due to their label-free real-time sensing capabilities. 
However, most of these nanosensors require functionalization for specific ion types \cite{ahoulou2021functionalization,ren2017nanopore,yao2021carbon}. 
Universal ion sensors that are sufficiently sensitive to detect individual ions have yet to be developed.

In this paper, we propose a concept for an ultrasensitive single-ion detector based on carbon nanotube field-effect transistors (CNT FETs). 
CNTs have traditionally been proposed as chemical sensors by detecting change in conductivity through surface reaction and adsorption~\cite{schroeder2018carbon, norizan2020carbon,acharya2024single}. 
In contrast, we use CNTs as ion channels that confine ion propagation to the 1D CNT interior but allow for continuous propagation of the ion without physisorption or chemisorption of the ion on the CNT surface.
In consequence, the CNT ion sensors allow for non-destructive ion sensing. 
Any ion that enters the CNT will be detected, since no ion-specific functionalization is needed. 
The CNT sensor can detect ions of both polarities and is compatible with nanodevice typical signal-improving electronics~\cite{shim2007stochastic}. 
%It is also possible to do selective ion detection, if needed, by functionalizing the entry of the CNT \cite{samoylova2017selective}.

\begin{figure*}
  \centering
  \includegraphics[width=6.69in]{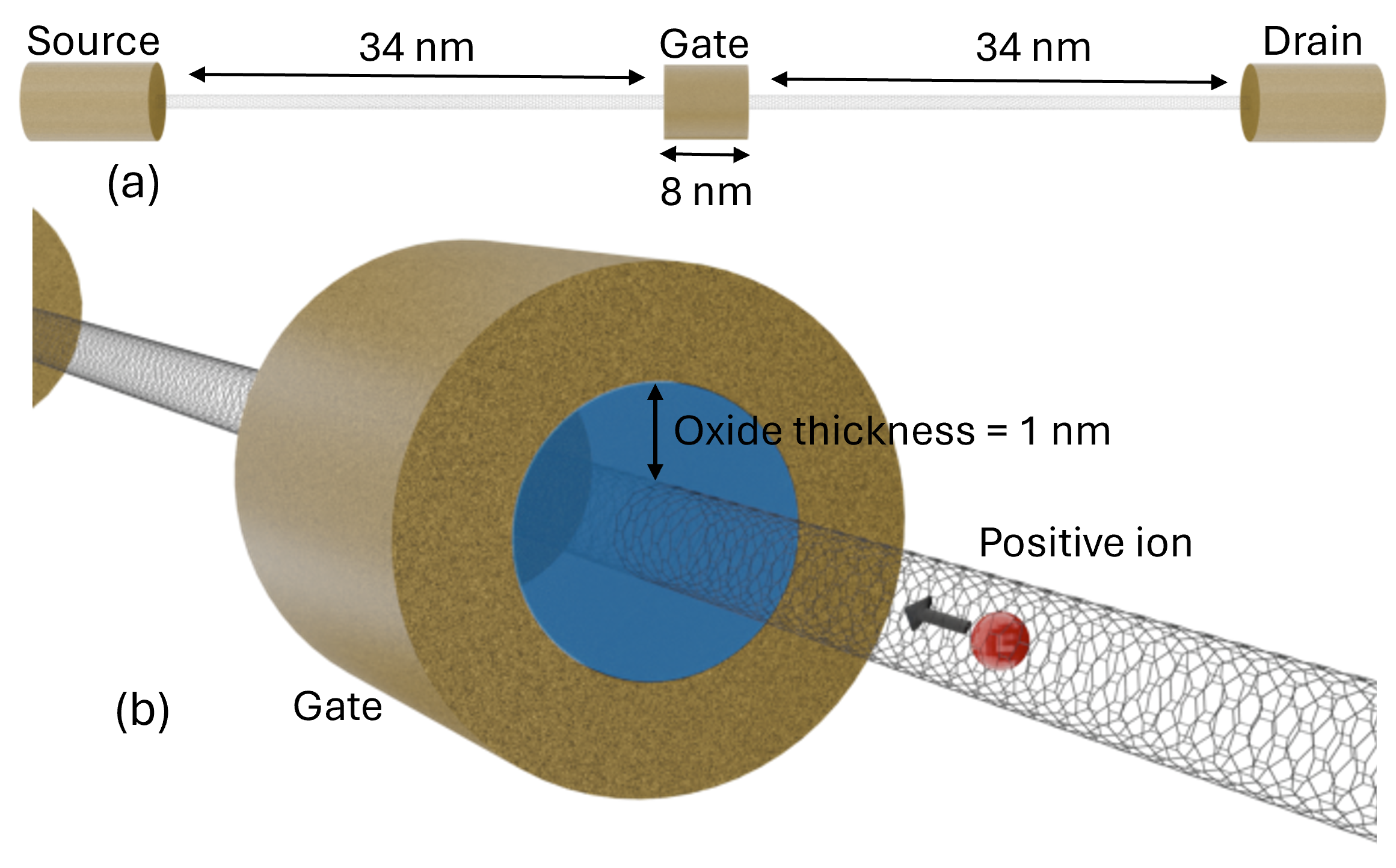}
  \caption{Schematic of the CNT based single-ion detector: The side view in (a) shows the dimensions of the specific CNT FET structure considered in the main text. Ions within the CNT channel get detected when they enter the gate region as illustrated in (b). }
  \label{fig1}
\end{figure*}

The core of the proposed device is a semiconducting CNT FET structure~\cite{franklin2013carbon,lin2023scaling} in Figure~\ref{fig1} that detects ions that propagate through the CNT channel. 
%The CNT FET acts as a nanoscale ion charge amplifier: the ion modifies the drain-source current (Ids) of the CNT FET when it passes under the gate.
%The change in current by the passing ion can be several orders of magnitude when a suitable operating point such as the subthreshold regime is selected. 
Once the ion is sufficiently close to the gate region, it transforms the CNT FET into a resonant tunneling diode (RTD)~\cite{bowen1997quantitative}: 
The attractive charge of the ion yields discrete quantum dot states in the CNT. 
Electrons or holes that have to pass the barrier in the gate region can use these quantum dot states for resonance tunneling through the barrier.
%When the ion is in the gate barrier region, the barrier region is thin enough to allow tunneling through the ion bound energy levels.
This process changes the source-drain current ($I_{ds}$) of the CNT device by several orders of magnitude when the gate voltage is tuned to maximize the resonant tunneling. 
For positive (negative) ions, the CNT FET will be configured for n-type (p-type) operation.
%The CNT FET is embedded into a dielectric material that prevents electron leakage out of the CNT (Refs). 
The channel gate electrode is a gate-all-around electrode~\cite{franklin2013carbon} for maximal ion sensitivity.
Back gate electrodes~\cite{chiu2025overcoming} are used to control the effective doping density and doping type.
Electric fields at the entry of the CNT can funnel ions into the CNT channel~\cite{su2011control} if needed.

\subsection{Methodology}
All electronic transport results presented in this paper are solved in the Nonequilibrium Green’s Function (NEGF) implementation of the nanodevice simulation tool NEMO5~\cite{steiger2011nemo5}. 
Electrons are discretized in the atomic representation of Ref.~\citenum{neophytou2007non}. 
Electronic energies in the Dyson and Keldysh equations are self-adaptively discretized to reliably resolve ion-induced electronic resonances and band edges of the CNT-FET~\cite{eiseman1987adaptive}.
Charge self-consistency is ensured by iterating the NEGF equations with the nonlinear Poisson equation~\cite{birner2007nextnano}. 
The Poisson equation is solved on the finite element mesh of libMesh~\cite{kirk2006libmesh} which was adapted to the respective atomic geometry of the CNT and the ion in all three dimensions.
Source and drain contacts are treated as Ohmic, while the gate contact is included as a Dirichlet boundary condition with cylindrical symmetry around the CNT. 
The charge distribution of all electrons, including deep-lying valence band electrons and electrons of partly filled conduction bands, is solved explicitly within the NEGF approach of Ref.~\citenum{chu2018explicit}.
In this way, the electrostatic screening of valence band electrons that do not take part in transport is explicitly included for every voltage and ion location.
This lifts the requirement of defining highly anisotropic dielectric screening constants within and perpendicular to the CNT surface. 
Correspondingly, the dielectric constant in the Poisson equation is set to the one of vacuum. 
%The Poisson equation is solved with a finite elements mesh. FEM mesh and adaptive energy resolution are increased until an excellent overall convergence is achieved for every bias point and ion position. 
Ions are represented as point charges in the Poisson equation that interact with CNT electrons only electrostatically.
%This is justified with DFTB calculations showing that there is no significant charge redistribution between the ion and CNT as discussed in the appendix A1 (do we have that appendix? .

\subsection{Results and discussion}
\subsubsection{Sensor operating principle}
The sensor concept is exemplified with a n-type semiconducting (11,0) single-walled CNT FET (diameter of 8.7~\AA~) at room temperature. 
76~nm of the CNT FET are explicitly solved with NEGF, while the remaining homogeneous CNT is covered with open boundary conditions.
An 8~nm long cylindrical gate is centered around a 8~nm long intrinsic CNT section that is surrounded by 34~nm long source and drain regions with an n-type doping density of $1.5~\times 10^{19} \,\text{cm}^{-3}$. 
This doping density corresponds to an electron density of 0.16~$e^-$/nm, which is within the experimentally measured range~\cite{leecompact}.
The drain-source voltage ($V_{ds}$) is set to 0.05~V.

\begin{figure}
  \centering
  \includegraphics[width=3.37in]{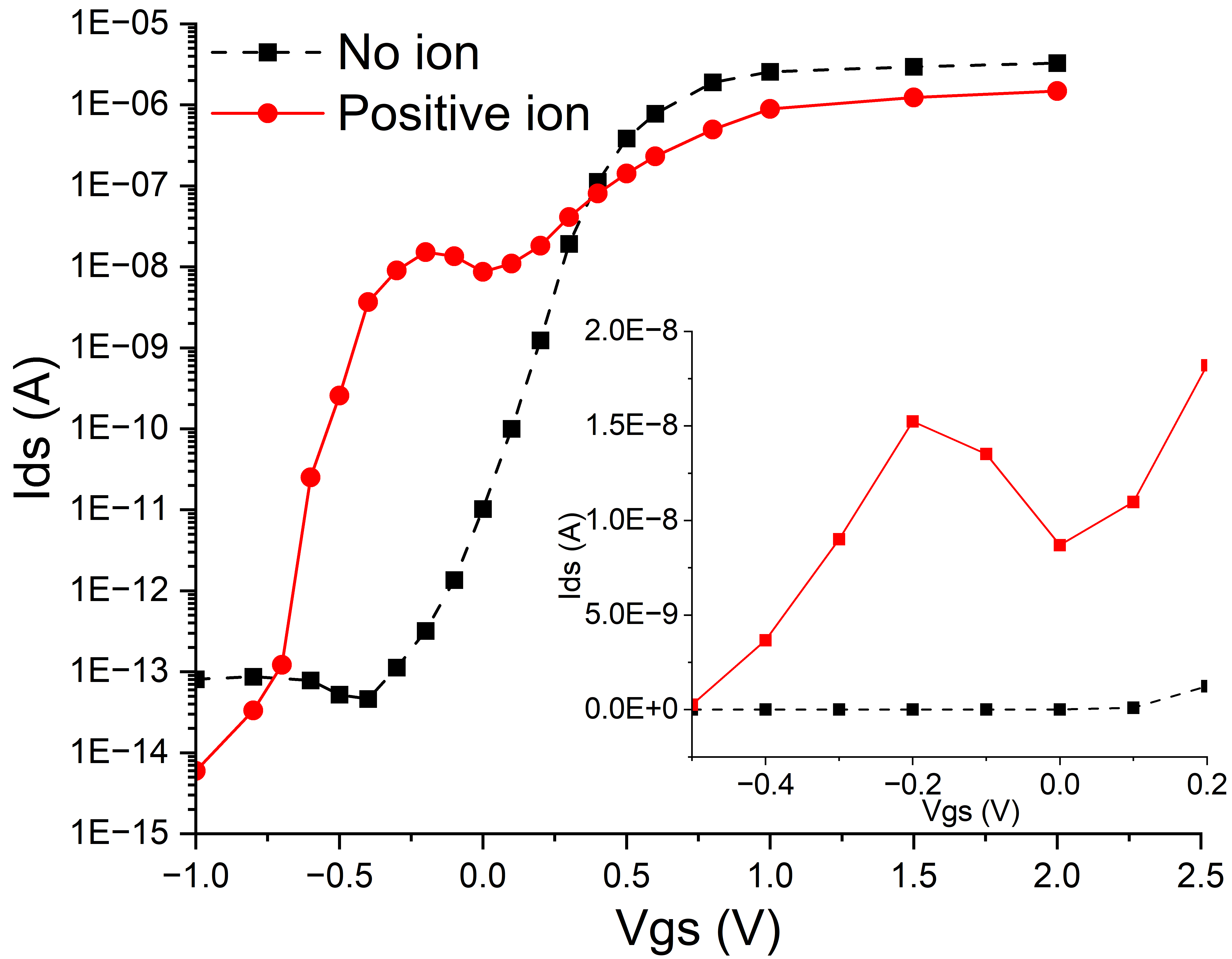}
  \caption{Drain-source current density of the single ion detector of Fig.~\ref{fig1} with (red solid lines) and without (black dashed lines) a positive ion in the CNT channel under the gate. An ion with a charge of 1~e under the gate switches the device's performance from a standard FET characteristics to the one of a resonant tunneling diode. The negative differential resistance of the diode is highlighted in the linear-scale inset. The sensor's highest sensitivity is around the diode's current density peak at a gate bias of -0.2~V in the considered device.
  }
  \label{fig2}
\end{figure}

\begin{figure*}
  \centering
  \includegraphics[width=6.69in]{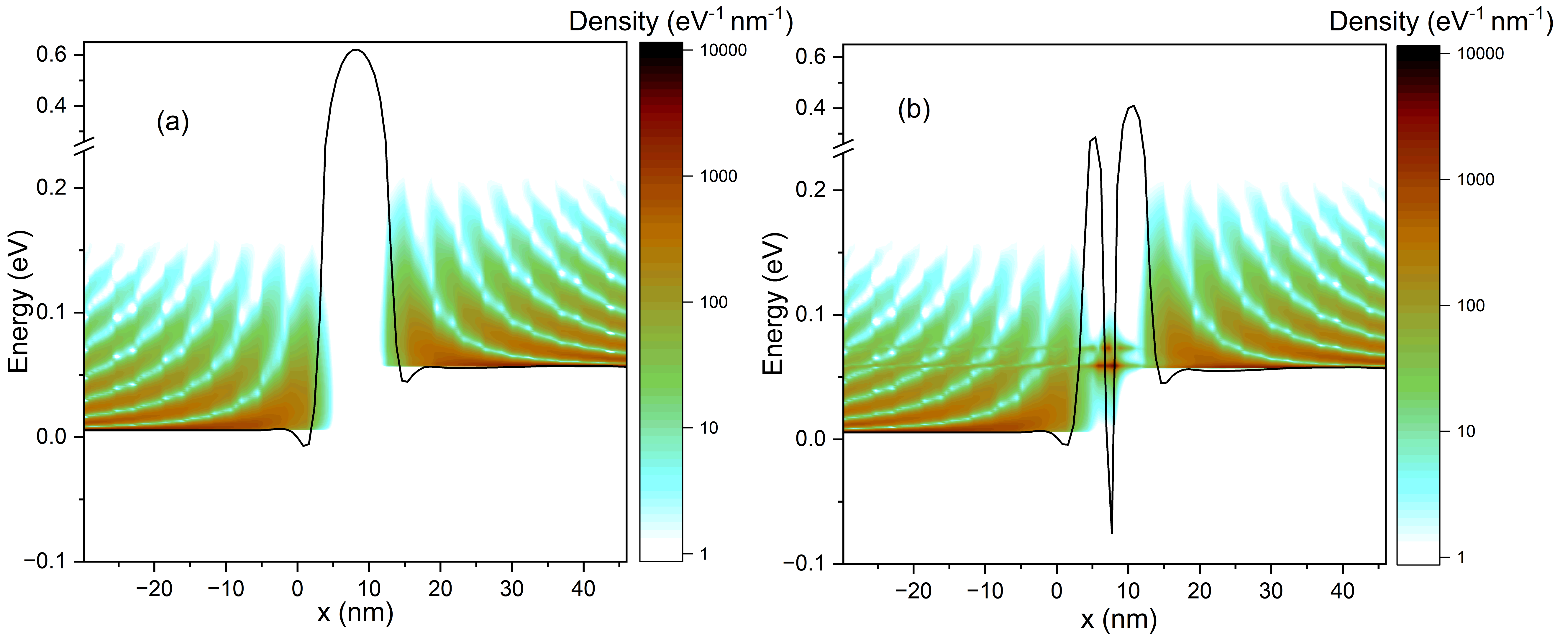}
  \caption{Conduction band energy (line) and contour graph of the spatially and energy resolved electron density along the surface of the CNT of Fig.~\ref{fig1} at the resonant gate voltage of -0.2~V for (a) no ion in the CNT and (b) a positive ion centered in the CNT channel under the gate at 7.5~nm. The gate extends from 4.2~nm to 12.2~nm.}
  \label{fig3}
\end{figure*}

Figure~\ref{fig2} shows the drain current, $I_{ds}$, as a function of the gate voltage ($V_{gs}$) for two device scenarios: 
No ion in the CNT FET (dashed black line) and a positive ion in the center of the gate region (solid red line). 
Comparison of the I-V characteristics demonstrates the positive ion in the CNT channel effectively transforms the CNT FET into a CNT-based RTD:
The I-V characteristic of the CNT FET with the ion has a pronounced negative differential resistance (NDR) right after the current maximum at -0.2~V gate voltage (highlighted by the inset in Fig.~\ref{fig2}).
The positive ion induces a quantum well with resonant energy levels, as illustrated in detail in Fig.~\ref{fig3}:

Figure~\ref{fig3} shows the contour graphs of the position and energy resolved electron density for the case with (a) no ion and with (b) a positive ion in the CNT channel at an applied gate voltage of -0.2~V. 
The solid line in both graphs indicates the respective effective conduction band edge as a function of position.
The conduction band edge in Fig.~\ref{fig3}(a) resembles the typical potential profile of n-type FETs with intrinsic gate regions~\cite{dang2006cntfet}.
At the given $V_{gs}$ of -0.2~V, the CNT FET is effectively turned off and electrons are completely reflected at the gate barrier.
This causes the interference pattern of the energy and position resolved density in Fig.~\ref{fig3}(a).
In contrast, the positive ion induces a quantum well in the middle of the gate barrier which yields a RTD-typical double barrier profile~\cite{bowen1997quantitative}. 
The maxima of the spectral function in the gate region (between 4.2~nm and 12.2~nm) in Fig.~\ref{fig3}(b) correspond to the quantum confined states of that induced quantum well.
For the shown $V_{gs}$ of -0.2~V, the first quantum confined state aligns with the conduction band edge of the CNT source side (at 50~meV).
This alignment allows electrons to resonantly tunnel through the gate region and thereby to maintain  $I_{ds}$ of nearly 5 orders of magnitude larger than the current in the ion-free CNT FET case. 
The linewidths of the involved quantum confined states determine the height and width of the RTD IV resonance~\cite{bowen1997quantitative}.

In consequence, the presented single ion sensor operates with the highest sensitivity when its gate voltage is set to the resonant voltage of the CNT RTD (here at $V_gs$=-0.2V):
When no ion is in the channel, the sensor is turned off and $I_{ds}$ vanishes. 
With an ion in the gate region, the sensor turns into a CNT RTD at its resonance condition i.e., it maintains a high $I_{ds}$.

%This is discussed in more detail in the supplements (see Figs.~S1 and ~S2).

\begin{figure}
  \centering
  \includegraphics[width=3.37in]{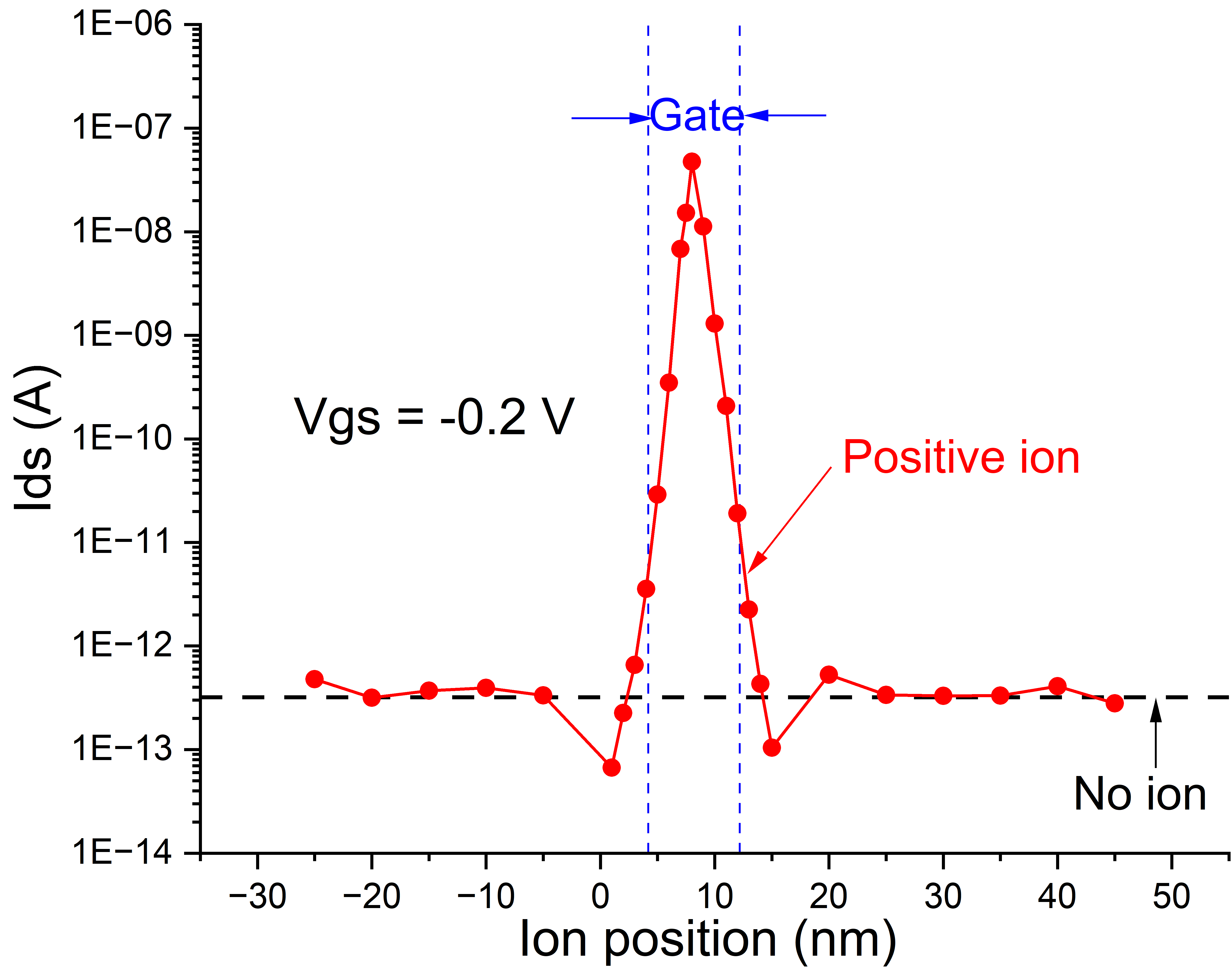}
  \caption{Source-drain current density of the CNT sensor of Fig.~\ref{fig1} as a function of the ion position centered in the channel along the CNT axis for a gate voltage of -0.2~V. The solid line is only meant to guide the eye. The dashed horizontal line shows the drain current without an ion in the channel. The dashed vertical lines indicate the gate region.}
  \label{fig4}
\end{figure}

Figure~\ref{fig4} shows $I_{ds}$ as a function of the ion position along the CNT axis.
As can be seen from the graph, the ion induces a source-drain current only when it is within the gate region (marked by vertical lines).

\begin{figure*}
  \centering
  \includegraphics[width=6.69in]{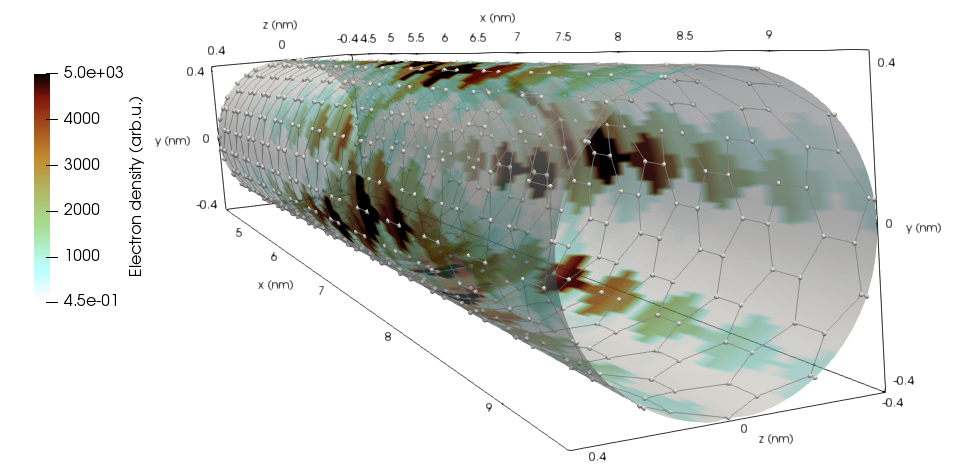}
  \caption{Contour graph of the electron density of the first ion-induced quantum confined state at $E=0.059~\textit{eV}$ of Fig.~\ref{fig3}(b) projected on the CNT surface. The positive ion is placed at the CNT center axis at x=7.5~nm.}
  \label{fig5}
\end{figure*}

The ion in the center of the gate region induces quantum confined states that wrap around the complete CNT circumference as can be seen in Fig.~\ref{fig5} for the first quantum confined state at the energy of $E=0.059~\textit{eV}$.

\subsubsection{Ion charge state detection}
\begin{figure}
  \centering
  \includegraphics[width=3.37in]{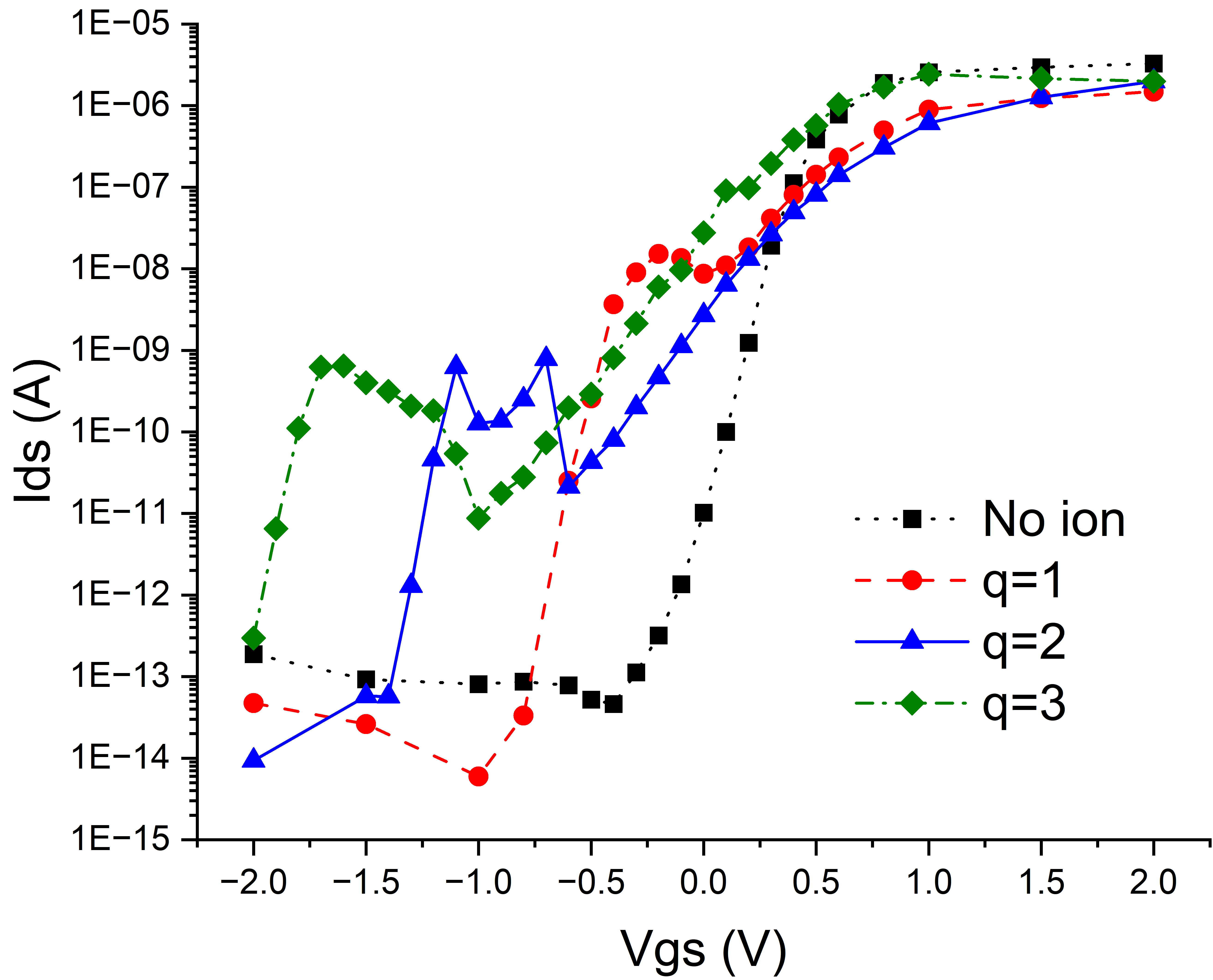}
  \caption{Drain-source current density of the CNT sensor for the cases of no ion (black) and with ions of charge states q=1 (red), q=2 (blue), and q=3 (green).
  }
  \label{fig6}
\end{figure}

The CNT sensor is sensitive to the charge state of the ion. 
Figure~\ref{fig6} compares the I-V characteristics of the CNT sensor without any ion in the nanochannel with the I-V characteristics of the CNT sensor with ions of various charge states (q=1, q=2 and q=3).
For q=2 and q=3, the ions were modeled as two and three point charges, respectively, separated by 10~\AA~.
With different ion charges, the electrostatically induced quantum confinement on the CNT surface changes.
This in turn shifts the resonant condition that drives the transition from CNT FET to RTD and consequently modifies the I-V characteristics (see Fig.~\ref{fig6}). 
Overall, each of the 4 ion configurations (q=0,1,2,3) can be distinguished with the drain-source current for a given gate voltage $V_{gs}$.
Higher charged ions can induce more quantum confined states on the CNT surface which yield I-V characteristics with multiple resonances.  

%For q=2 and q=3, the resonance occurs in the OFF regime of the CNT FET, though this operating point can be tuned through CNT device design. 
%The RTD response for the higher charge states also shows two distinct peaks, which arises from the alignment of two quantum-confined states with the conduction-band edge.
%The separation between these states determines the spacing and width of the resonant features.

%The ion sensor's operating principles have been confirmed for CNT FETs of different chiralities, larger diameters, and gate lengths.
%Simulations with larger diameter CNTs and longer gate lengths show the same effect of an ion in the gate region transforming the CNT FET into a RTD. This confirms that the operating principle of the CNT ion sensor is not limited by the diameter/chirality of the CNT or gate lengths.

\subsubsection{CNT sensor design aspects}
The presented ion detector's sensitivity can be optimized with device geometry aspects such as the CNT diameter and gate length.
Figure~S1 shows the CNT ion detector for 3 different gate lengths. 
The gate length shifts the I-V characteristics and modifies the gate voltage at which the sensor is most responsive i.e. at which gate voltage the ion-induced transition to RTD has largest impact on the drain-source current.

Larger CNT diameters can become necessary when the molecules under consideration exceed the CNT diameter. 
It can be expected that larger distances of charges to the CNT surface reduce the induced Coulomb potential and thereby the detector's signal strength.
Figure~S2 demonstrates that the ion-induced FET-to-RTD transition remains when the CNT diameter is significantly increased:
It shows the I-V characteristics of a (19,0) CNT single ion detector i.e. of a CNT with diameter of 1.5~nm which is ~2x the diameter of the (11,0) CNT considered in Figs.~2-5.
The switching mechanism of FET to RTD and a signal strength of 400x can be observed.
Optimization of the design details of the (19,0) CNT ion detector can be expected to increase that signal strength significantly.
Similar to the impact reduction by distance between ion and CNT, solvents surrounding the ion can partially screen its induced electrostatic impact on the CNT. 
This depends on the solvent's dielectric constant and does not alter the detector's fundamental functionality.
Figure~S3 confirms the operation of the detector in p-type mode for negatively charged ions.

This work is purely computational; experimental studies are needed to establish the measurable response of the sensor under realistic operating conditions. Several ion systems are suitable candidates for such studies. Transport of K$^+$ and Cl$^-$ through individual CNTs with a diameter of 0.8 nm, comparable to the (11,0) CNT considered here, has been experimentally demonstrated \cite{tunuguntla2017enhanced}. Divalent Ca$^{2+}$ has been studied in CNT porins of similar diameter \cite{park2023probing}. Trivalent rare-earth ions have been transported through 1.5 nm diameter CNT porins \cite{abdullah2026rare}, which matches the diameter of the (19,0) CNT detector in Fig. S2.
%To assess the practical viability of the proposed CNT sensor, we examine its robustness under several variations in device conditions, including gate length, CNT diameter, and pFET-based configurations.
%These cases are not individually optimized to maximize performance; rather, they are included to show that the RTD-based sensing principle remains valid across different device scenarios.

%The sensor performance remains robust against variations in gate length, as shown in Fig. S1. Although changing the gate length from 6 nm to 10 nm shifts the transfer characteristics and modifies the exact gate voltage at which the ion-induced transition occurs, the key sensing behavior is preserved in all cases: the device still switches from a low-current FET-like state to a higher-current RTD regime upon ion presence.

%A larger diameter (19,0) CNT is considered to verify that the RTD behavior is preserved beyond the smaller-diameter case. Fig S2 demonstrates that the ion-induced FET-to-RTD transition remains observable when the CNT diameter is increased. This supports the generality of the sensing mechanism and shows that the RTD principle is not limited to a single CNT chirality.

%Finally, we evaluate the sensing mechanism using a p-type CNT FET to demonstrate that the proposed detector can work for negative ions. As shown in Fig. S3, the sensor exhibits the expected FET to RTD behavior with the presence of a single negative ion.

\subsection{Conclusion}
In summary, this paper introduces and quantitatively assesses a new type of CNT based single-ion detector.
Unlike conventional CNT sensors, the proposed device does not need functionalization and can serve as a universal ion detector.
The sensor behaves like a CNT FET when no ion is in the CNT channel. 
The Coulomb potential of a single ion in the CNT channel under the gate region changes the electronic device behavior from FET into RTD.
Once the ion leaves the gate region the sensor returns to the FET behavior.
The concrete sensor example of this work was predicted to show a signal-to-noise ratio of 5 orders of magnitude. 
Such an ultrasensitive single-ion sensor can enable breakthroughs in biomedical, environmental, and quantum applications.

\begin{acknowledgments}
Computational resources were provided by the Negishi Research Computing Cluster (RCAC) at Purdue University. The authors also thank Manas Pratap for his assistance in creating the figures and improving the visual quality of this work.
\end{acknowledgments}

\section*{Author Declarations}
\subsection*{Conflict of Interest}
The authors have no conflicts to disclose.

%\subsection*{Author Contributions}
%\textbf{Namita Narendra}: Conceptualization (supporting); Investigation; Writing - original draft. \textbf{Tillmann Kubis}: Conceptualization (lead); Supervision; Writing - review and editing. 

\section*{Data Availability}
The data that support the findings of this study are available from the corresponding author upon reasonable request.
\bibliography{CNT_ion_sensor.bib}

\end{document}